\documentclass[11pt]{article}

\usepackage[a4paper,margin=1in]{geometry}
\usepackage{amsmath,amssymb}
\usepackage{booktabs}
\usepackage{microtype}
\usepackage[american]{babel}
\usepackage{hyperref}
\usepackage{xcolor}
\usepackage{graphicx}
\usepackage{upgreek}
\usepackage{caption}
\hypersetup{
  pdflang={en-US},
  colorlinks=true,
  linkcolor=blue!50!black,
  citecolor=blue!50!black,
  urlcolor=blue!50!black,
  pdftitle={Axion Minihalo Disruption in the Milky Way: Effects of Disk and Bulge Gravity},
  pdfauthor={Chris Gordon and Marc Kamionkowski}
}

\newcommand{\Msun}{M_{\odot}}
\newcommand{\pc}{\mathrm{pc}}
\newcommand{\kpc}{\mathrm{kpc}}
\newcommand{\kms}{\mathrm{km\,s^{-1}}}

\title{Axion Minihalo Disruption in the Milky Way: Effects of Disk and Bulge Gravity}
\author{
Chris Gordon$^{1}$ and Marc Kamionkowski$^{2}$\\[0.5em]
\small $^{1}$School of Physical and Chemical Sciences, University of Canterbury\\
\small Christchurch 8140, New Zealand\\
\small $^{2}$The William H. Miller III Department of Physics and Astronomy, Johns Hopkins University\\
\small Baltimore, Maryland 21218, USA
}
\date{}

\begin{document}
\maketitle

\begin{abstract}
In post-inflationary axion scenarios, primordial density fluctuations form miniclusters that subsequently merge into larger minihalos.
The disruption of axion minihalos by stellar encounters depends on their Galactic orbits. We extend a successive-encounter calculation by including the gravitational potentials of the stellar disk and bulge, as well as encounters with bulge stars. For a smoothly truncated Navarro--Frenk--White Galactic dark-matter halo, we reconstruct local equilibrium velocity distributions using Schwarzschild orbit superposition and evaluate the resulting stellar-induced minihalo disruption. At $m_a=25\,\upmu\mathrm{eV}$, the surviving mass fraction above $10^{-12}\,\Msun$ is $(33.23\pm0.05)\%$ when only dark-matter gravity and disk stellar encounters are included, $(28.31\pm0.08)\%$ when disk gravity is also included, and $(28.06\pm0.02)\%$ when bulge gravity and bulge stellar encounters are added. Together, disk gravity, bulge gravity, and bulge stellar encounters reduce minihalo survival by $(5.17\pm0.05)$ percentage points. The quoted uncertainties reflect sensitivity to the size of the orbit library. Motivated by the fiducial model's relatively high circular speed of $249\,\kms$ at $8\,\kpc$ and independent estimates of the stellar disk mass, we also reduce the disk mass from $5.76\times10^{10}\,\Msun$ to $5.00\times10^{10}\,\Msun$. This lowers the circular speed to $241\,\kms$ and increases survival to $(29.70\pm0.12)\%$. 

\end{abstract}

\section{Introduction}

The Peccei--Quinn mechanism was proposed to resolve the strong-CP problem in quantum chromodynamics (QCD)~\cite{PecceiQuinn1977}. It predicts a light pseudoscalar, the axion~\cite{Weinberg1978,Wilczek1978}, which is also a well-motivated dark-matter candidate~\cite{PreskillWiseWilczek1983,AbbottSikivie1983,DiLuzio2020}. If Peccei--Quinn symmetry is broken after inflation, the axion field has different initial values in causally disconnected regions. Nonlinear field dynamics around the QCD epoch then generate substantial small-scale density fluctuations, which collapse gravitationally into axion miniclusters. Subsequent hierarchical merging produces larger minicluster halos, or minihalos, whose formation, mass functions, and internal structures have been explored analytically and numerically (see, e.g., Refs.~\cite{HoganRees1988,KolbTkachev1993,Eggemeier2020,Xiao2021}). Their present-day abundance depends not only on their formation in the early Universe, but also on their subsequent evolution within the Milky Way.

If a large fraction of axion dark matter remains bound in compact miniclusters, the density encountered by a terrestrial haloscope need not equal the smooth local density inferred from stellar dynamics. Eggemeier et al.~\cite{Eggemeier2023} studied the underdense regions between axion miniclusters, or minivoids, and their consequences for direct detection. Material stripped from miniclusters by stellar encounters is redistributed into tidal streams. O'Hare et al.~\cite{OHare2024} modeled this debris in the solar neighborhood and showed how overlapping streams could replenish the local axion density and produce narrow spectral features in haloscope searches. O'Hare and Pierobon~\cite{OHarePierobon2026} explored the effect of fine-grained velocity substructure on high-resolution axion haloscopes.

In the Milky Way, encounters with individual stars inject energy into minihalos, potentially removing part or all of their bound mass. The resulting disruption depends on the minihalo's internal density profile, the sequence of encounters, and its Galactic orbit. 
Kavanagh et al.~\cite{Kavanagh2021} investigated stellar disruption using Monte Carlo simulations of successive encounters with stars in the Galactic disk and bulge, tracking mass loss. They approximated minicluster orbits as Keplerian ellipses about the Galactic center, taking the effective central mass for each orbit to be the mass enclosed within its semimajor axis in a spherical Navarro--Frenk--White (NFW) model of the Galactic dark-matter halo~\cite{NavarroFrenkWhite1996}. The gravitational contributions of the stellar disk and bulge were not included in the orbital calculation. 

Shen et al.~\cite{Shen2024} subsequently developed a framework combining the hierarchical assembly and infall of dense dark-matter minihalos with stellar encounters and smooth tidal stripping in the Milky Way. They found that stellar encounters dominated minihalo disruption, while smooth tidal stripping by the Galactic potential had a relatively small effect on the resulting mass function. They calibrated their treatment of individual stellar encounters and tidal effects using idealized $N$-body simulations and examined several uncertainties in the Galactic environment.  Their orbital averaging employed a simplified spherical isothermal description. Shen et al.\ modeled repeated stellar encounters by accumulating the injected energy and applying a mass-loss prescription calibrated to individual encounters. For their fiducial axion-minicluster model, with a minimum adiabatic host-halo mass of $10^{2}\,\Msun$,\footnote{Shen et al.\ state a value of $10^{-2}\,\Msun$, but their calculation used $10^{2}\,\Msun$~\cite{DSouza2025}.} they found a surviving mass fraction of approximately $58\%$ above $10^{-12}\,\Msun$ at the solar radius.

A further issue is the combined effect of multiple stellar encounters. DSouza and Gordon~\cite{DSouzaGordon2024} studied the response of minihalos to successive encounters and showed that adding the associated energy injections linearly can underestimate mass loss when a minihalo has time to relax between perturbations. Building on this result, DSouza et al.~\cite{DSouza2025} introduced a hybrid prescription that treats closely spaced encounters differently from encounters separated by a minihalo dynamical time. For an axion mass $m_a=25\,\upmu\mathrm{eV}$ and a minimum adiabatic host-halo mass of $10^2\,\Msun$, this reduced the surviving minihalo mass fraction above $10^{-12}\,\Msun$ from $53.5\%$ for linear energy addition to $29.6\%$ with the hybrid prescription. In that calculation, Galactic trajectories were generated in a spherical singular-isothermal potential, while the stellar disk supplied the disrupting encounters. The gravitational effects of the disk and bulge on the equilibrium minihalo velocities and on those trajectories were therefore not included in the model.

 O'Hare et al.~\cite{OHare2024} integrated minicluster trajectories in a multicomponent Galactic potential but drew initial velocities from a Maxwellian halo model. Here we reconstruct the local velocity distribution for each prescribed potential before evaluating the disruption histories.

For a spherical halo with an isotropic velocity distribution, the phase-space distribution can be constructed using Eddington inversion~\cite{Eddington1916}. We use this as an independent benchmark for the spherical truncated-NFW Schwarzschild calculation, which serves as the baseline for comparisons with baryonic gravity. Here a benchmark means a separate calculation used to check our method. The Eddington benchmark uses the same potential as the spherical Schwarzschild calculation but a different construction of the velocity distribution. Once the disk potential is included, this isotropic construction is no longer sufficient. To treat all Galactic configurations consistently, we use the Schwarzschild orbit-superposition method~\cite{Schwarzschild1979}, in which non-negative weights are assigned to orbits in a prescribed gravitational potential to reproduce a target dark-matter density~\cite{BinneyTremaine2008,VasilievAthanassoula2015}. We implement the Galactic potentials and orbital calculations using AGAMA~\cite{Vasiliev2019}. Related studies have constructed Galactic dark-matter velocity distributions in baryonic potentials using spherical inversion methods or axisymmetric distribution functions~\cite{CatenaUllio2012,LacroixStrefLavalle2018,PetacUllio2019}.

We retain the minihalo population and hybrid encounter prescription of DSouza et al.~\cite{DSouza2025} but replace their singular-isothermal Galactic halo with a smoothly truncated NFW profile. As a second benchmark, we repeat their singular-isothermal calculation using AGAMA to check our implementation. We compare halo-only, halo-plus-disk, and halo-plus-disk-plus-bulge models. We also test a lower stellar disk mass while holding the other components fixed. 

\section{Methods}
\label{sec:methods}

\subsection{Galactic mass model}
\label{subsec:galactic-mass-model}

We extend the calculation of DSouza et al.~\cite{DSouza2025} by including the gravitational effects of the Milky Way stellar disk and bulge when determining the initial minihalo velocities and their subsequent Galactic orbits. We prescribe a spherical dark-matter density throughout, including in the models with a stellar disk and bulge. The halo follows a smoothly truncated NFW (tNFW) profile in terms of the spherical Galactocentric radius $r$. We define the NFW density by
\begin{equation}
\rho_{\rm NFW}(r)
=
\frac{\rho_s}{(r/r_s)(1+r/r_s)^2},
\label{eq:tnfw-inner}
\end{equation}
with
\begin{equation}
\rho_s=0.014\,\Msun\,\pc^{-3},
\qquad
r_s=16.1\,\kpc.
\label{eq:tnfw-params}
\end{equation}
Here $\rho_s$ and $r_s$ are the NFW characteristic density and scale radius, respectively. We adopt a truncation radius $r_{\rm cut}=161\,\kpc=10r_s$ and, following the exponentially truncated NFW prescription of Springel and White~\cite{SpringelWhite1999} (see also Ref.~\cite{Drakos2017}), define the full halo density as
\begin{equation}
\rho_{\rm DM}(r)
=
\begin{cases}
\rho_{\rm NFW}(r), & r\leq r_{\rm cut},\\[4pt]
\rho_{\rm NFW}(r_{\rm cut})
\left(\dfrac{r}{r_{\rm cut}}\right)^{\delta}
\exp\left[-\dfrac{r-r_{\rm cut}}{r_{\rm decay}}\right],
& r>r_{\rm cut},
\end{cases}
\label{eq:tnfw-outer}
\end{equation}
where $r_{\rm decay}=161\,\kpc$ and
\begin{equation}
\delta=
\frac{r_{\rm cut}}{r_{\rm decay}}
-
\frac{1+3x_{\rm cut}}{1+x_{\rm cut}},
\qquad
x_{\rm cut}=\frac{r_{\rm cut}}{r_s},
\label{eq:tnfw-delta}
\end{equation}
so that the density and its first derivative are continuous at $r_{\rm cut}$. The inner NFW parameters coincide with those of the NFW Milky Way mass-model fit of Nesti and Salucci~\cite{NestiSalucci2013}. 

For the fiducial calculation, the stellar disk is modeled using the thin- and thick-disk density adopted by DSouza et al.~\cite{DSouza2025}, with surface-density normalizations and scale lengths from the best-fit disk model of McMillan~\cite{McMillan2011}, as used by Shen et al.~\cite{Shen2024},
\begin{equation}
\rho_{\rm disk}(R,z)
=
\sum_d
\frac{\Sigma_{d,0}}{2h_d}
\exp\left(-\frac{R}{R_d}-\frac{|z|}{h_d}\right).
\label{eq:disk-density}
\end{equation}
Here $R$ is the cylindrical Galactocentric radius, $z$ is the height above the Galactic midplane, and the index $d$ labels the thin- and thick-disk components. The parameters $\Sigma_{d,0}$, $R_d$, and $h_d$ are the central surface density, radial scale length, and vertical scale height of component $d$, respectively.
For the thin disk we use
\begin{equation}
(\Sigma_{d,0},R_d,h_d)
=
(816.6\,\Msun\,\pc^{-2},
 2.90\,\kpc,
 0.30\,\kpc),
\label{eq:thin-disk-params}
\end{equation}
and for the thick disk
\begin{equation}
(\Sigma_{d,0},R_d,h_d)
=
(209.5\,\Msun\,\pc^{-2},
 3.31\,\kpc,
 0.90\,\kpc).
\label{eq:thick-disk-params}
\end{equation}
The resulting total disk mass is $5.76\times10^{10}\,\Msun$.

For the bulge we use the axisymmetrized form of the model of Bissantz and Gerhard~\cite{BissantzGerhard2002} given by McMillan~\cite{McMillan2011} and adopted by Kavanagh et al.~\cite{Kavanagh2021},
\begin{equation}
\rho_{\rm bulge}(R,z)
=
\rho_0
\frac{\exp[-(r'/r_{\rm cut,b})^2]}
{(1+r'/r_{\rm b})^\alpha},
\label{eq:bulge-density}
\end{equation}
where
\begin{equation}
r'=\sqrt{R^2+(z/q)^2},
\label{eq:bulge-radius}
\end{equation}
with
\begin{equation}
\rho_0=99.3\,\Msun\,\pc^{-3},
\qquad
q=0.5,
\qquad
r_{\rm b}=0.075\,\kpc,
\label{eq:bulge-params-density-shape}
\end{equation}
\begin{equation}
\alpha=1.8,
\qquad
r_{\rm cut,b}=2.1\,\kpc.
\label{eq:bulge-params-cutoff}
\end{equation}
Here $\rho_0$ is the bulge density normalization, $q$ specifies the vertical flattening, $r_{\rm b}$ is the bulge scale radius, and $\alpha$ is the power-law exponent governing the density profile at intermediate radii.
This model has a total bulge mass $M_{\rm bulge}=8.96\times10^9\,\Msun$.

We implement the disk and bulge potentials with the AGAMA stellar-dynamics package~\cite{Vasiliev2019}. As a check on the resulting Galactic potential, the circular speed at $R=8\,\kpc$ is $168\,\kms$ for the dark-matter halo alone, $239\,\kms$ after including the disk, and $249\,\kms$ after including both disk and bulge.

The fiducial full-model circular speed is higher than representative estimates near the Sun, including $232.8\pm3.0\,\kms$ from McMillan~\cite{McMillan2017} and $236\pm7\,\kms$ from Reid et al.~\cite{Reid2019}. 
We retain the fiducial disk for a controlled comparison with the earlier disruption calculation.

As a mass-model sensitivity test, we also adopt a lower total stellar disk mass of $5.00\times10^{10}\,\Msun$, motivated by the disk mass in the NFW fit of Nesti and Salucci~\cite{NestiSalucci2013}. Their single exponential disk has a radial scale length of $3.1\,\kpc$, close to the $2.90\,\kpc$ and $3.31\,\kpc$ values adopted here for the thin and thick disks, respectively.
The adopted disk mass is also consistent with the independent estimate $(5.17\pm1.11)\times10^{10}\,\Msun$ of Licquia and Newman~\cite{LicquiaNewman2015}. We multiply both fiducial disk surface-density normalizations by the common factor $f_{\rm disk}=M_{\rm disk}/M_{\rm disk,fid}\simeq0.868$, where $M_{\rm disk,fid}=5.76\times10^{10}\,\Msun$ is the fiducial disk mass. This scaling retains their relative contributions, radial scale lengths, and scale heights. The same factor multiplies the stellar disk density used to compute disruptive encounters. The truncated-NFW halo and the bulge density, gravitational potential, and stellar-encounter contribution are unchanged. This gives a circular speed of $241\,\kms$ at $R=8\,\kpc$.

We consider the four Galactic models defined in Table~\ref{tab:model-definitions}, which distinguishes the gravitational components governing the minihalo trajectories from the stellar populations supplying disruptive encounters.

\begin{table}[htbp]
\centering
\caption{Definitions of the four Schwarzschild models. All contain the same spherical tNFW dark-matter halo. The fiducial and lighter stellar disks have masses of $5.76\times10^{10}\,\Msun$ and $5.00\times10^{10}\,\Msun$, respectively.}
\label{tab:model-definitions}
\begin{tabular}{@{}cp{0.41\textwidth}p{0.30\textwidth}@{}}
\toprule
Model & Gravitational potential & Stellar encounters \\
\midrule
1 & tNFW halo & Fiducial disk \\
2 & tNFW halo + fiducial disk & Fiducial disk \\
3 & tNFW halo + fiducial disk + bulge & Fiducial disk + bulge \\
4 & tNFW halo + lighter disk + bulge & Lighter disk + bulge \\
\bottomrule
\end{tabular}
\end{table}

The spherical Eddington calculation is a separate, unnumbered benchmark with the same gravitational potential and disk stellar encounters as Model~1 but a different construction of the initial-velocity distribution.

\subsection{Spherical velocity distributions}

For the singular-isothermal benchmark with constant circular speed $v_c=200\,\kms$, we draw initial velocities from the three-dimensional Maxwellian distribution~\cite{BinneyTremaine2008},
\begin{equation}
p_{\rm Max}(\mathbf v)
=
\frac{1}{(2\pi\sigma^2)^{3/2}}
\exp\!\left(-\frac{|\mathbf v|^2}{2\sigma^2}\right),
\qquad
\sigma=\frac{v_c}{\sqrt{2}},
\label{eq:maxwell-velocity-distribution}
\end{equation}
where $\sigma$ is the one-dimensional velocity dispersion.

For the independent spherical truncated-NFW Eddington benchmark, the isotropic phase-space distribution function $f_{\rm Edd}(\mathcal E)$ is obtained from Eddington inversion~\cite{Eddington1916,BinneyTremaine2008},
\begin{equation}
f_{\rm Edd}(\mathcal E)
=
\frac{1}{\sqrt{8}\,\pi^2}
\frac{d}{d\mathcal E}
\int_0^{\mathcal E}
\frac{d\Psi}{\sqrt{\mathcal E-\Psi}}
\frac{d\rho_{\rm DM}}{d\Psi},
\label{eq:eddington-inversion}
\end{equation}
where \(\Phi(r)\) is the Newtonian gravitational potential per unit mass of the truncated-NFW halo, \(\Psi(r)=\Phi(\infty)-\Phi(r)\) is the relative gravitational potential, \(\mathcal E=\Psi(r)-|\mathbf v|^2/2\) is the relative energy per unit mass, and \(\rho_{\rm DM}\) is regarded as a function of \(\Psi\). The normalized local three-dimensional velocity distribution is
\begin{equation}
p_{\rm Edd}(\mathbf v\mid r)
=
\begin{cases}
\dfrac{f_{\rm Edd}\!\left(\Psi(r)-|\mathbf v|^2/2\right)}
{\rho_{\rm DM}(r)}, & |\mathbf v|<\sqrt{2\Psi(r)},\\[8pt]
0, & \text{otherwise}.
\end{cases}
\label{eq:eddington-local-velocity-distribution}
\end{equation}
Both $p_{\rm Max}$ and $p_{\rm Edd}(\mathbf v\mid r)$ are normalized with respect to $d^3v$. For the Eddington benchmark, we draw velocities at $r=8\,\kpc$ from Eq.~\eqref{eq:eddington-local-velocity-distribution}.

\subsection{Schwarzschild construction of the dark-matter distribution function}

Once the disk is included, the Galactic potential is no longer spherical, so the isotropic Eddington construction used for the spherical benchmark is not applicable. We therefore use the Schwarzschild orbit-superposition method to approximate a stationary dark-matter distribution in each prescribed static total potential, following the general approach described in Sec.~4.7.2 of Binney and Tremaine~\cite{BinneyTremaine2008} and Vasiliev and Athanassoula~\cite{VasilievAthanassoula2015}. The non-negative weights of orbits integrated in that potential are fitted to the imposed spherical halo density on the finite spatial grid. This does not model the dynamical response of the halo to the assembly of the baryonic components.

For the main Schwarzschild results, including Model~1, we use $8\times10^5$ trial dark-matter orbits. We compare these solutions with $4\times10^5$-orbit reconstructions for all four models. For Model~4, additional calculations use $2\times10^5$ and $1.6\times10^6$ orbits. Initial positions are drawn from the target truncated-NFW density. Trial velocities are generated from an isotropic Eddington distribution constructed using the spherical monopole of the total potential. This proposal is used only to obtain phase-space coverage. In axisymmetric models, the library trajectories are integrated in the full axisymmetric potential.

Each orbit is followed for $50T_{\rm circ}$, where $T_{\rm circ}$ is evaluated with AGAMA in the spherically averaged total gravitational potential used to generate trial velocities and is the period of the circular orbit having the same energy as the trial orbit's initial phase-space point~\cite{Vasiliev2019}. For the axisymmetric models, the target dark-matter density is divided into cylindrical cells labeled by $(R,|z|)$. Each cell spans the full azimuthal range $0\leq\phi<2\pi$, and the use of $|z|$ applies the same spatial binning above and below the Galactic midplane. For each orbit, we record the fraction of its integration time spent in each cell. We use a nonuniform grid with 24 radial bins in \(R\) and 15 vertical bins in \(|z|\), with finer radial spacing around the adopted solar Galactocentric radius \(R_0=8\,\mathrm{kpc}\). The bin boundaries  are
\begin{align*}
R_{\rm edges}/\kpc ={}&
\{0,0.5,1,2,3,4,6,7.5,7.8,8,8.2,8.5,10,\\
&13,18,25,40,65,100,161,250,400,650,1000,1500\},\\
|z|_{\rm edges}/\kpc ={}&
\{0,0.15,0.3,0.6,1,2,4,8,16,32,64,128,\\
&256,512,1000,1500\}.
\end{align*}
The nonuniform spacing concentrates density constraints in the inner Galaxy and around the solar radius, where we sample the local dark-matter velocities. In particular, radial edges at $R=7.8$, $8.0$, and $8.2\,\kpc$ provide finer resolution near $R_0$, while the small-$|z|$ edges at $0.15$, $0.30$, $0.60$, and $1\,\kpc$ cover heights comparable to the adopted thin- and thick-disk scale heights of $0.30$ and $0.90\,\kpc$. Farther from the Galactic center and midplane, broader cells limit the number of sparsely populated density constraints. The radial edge at $161\,\kpc$ matches the halo's spherical truncation scale in the midplane, and the grid extends beyond this radius to cover the exponential outer tail. These boundaries are a practical choice guided by the model's physical scales, not a separately optimized spatial discretization.
Of the 360 cells, 328 are included in the density fit. We exclude cells containing less than $10^{-6}$ of the total target dark-matter mass. The excluded cells, together with the mass outside the grid, account for less than $0.1\%$ of the total dark-matter mass. The same grid and selection criterion are used for all axisymmetric models. For Model~1, the density constraints are instead imposed on a one-dimensional radial grid, as detailed in Appendix~\ref{app:spherical-schwarzschild}. We then determine non-negative orbit weights $w_i$ such that the weighted superposition of orbit densities reproduces the prescribed spherical dark-matter density. Denoting the weighted, time-averaged density contribution of orbit $i$ by $\rho_i(R,z)$, so that $w_i$ is already included in $\rho_i$, the Schwarzschild construction therefore seeks weights satisfying
\begin{equation}
\sum_i \rho_i(R,z) \simeq \rho_{\rm DM}(R,z).
\label{eq:schwarzschild-density-match}
\end{equation}
Here $\rho_{\rm DM}(R,z)=\rho_{\rm DM}(\sqrt{R^2+z^2})$ is the spherical target halo density expressed in cylindrical coordinates. For the axisymmetric models, the matching is imposed on the masses of the cylindrical grid cells used in the fit. For the optimization, we write the normalized orbit mass fractions as $a_i=w_i/\sum_k w_k$, with $a_i\geq0$ and $\sum_i a_i=1$. Let $p_{ij}$ denote the fraction of time orbit $i$ spends in fitted spatial cell $j$, $b_j$ the fraction of the total target dark-matter mass associated with that cell, $K$ the number of fitted constraints, and $N$ the number of trial orbits. We obtain the orbit weights by minimizing
\begin{equation}
Q=
\frac{1}{2K}
\sum_{j\in{\rm fit}}
\left[
\frac{\sum_i p_{ij}a_i-b_j}
{0.01\,b_j}
\right]^2
+
\frac{\lambda}{2N}
\sum_i
\left[
\frac{a_i-a_i^{\rm prior}}
{a_i^{\rm prior}}
\right]^2.
\label{eq:schwarzschild-objective}
\end{equation}
The factors of $1/2$ in the two quadratic terms are conventional normalizations, while the factors $1/K$ and $1/N$ average over fitted constraints and trial orbits. The factor $0.01$ sets a $1\%$ fractional density mismatch as the reference scale for the density term. The reference weights in the regularization term are equal, $a_i^{\rm prior}=1/N$. We use $\lambda=0.1$ for the main Schwarzschild comparisons and retain an unregularized ($\lambda=0$) fit to the same $4\times10^5$-orbit fiducial library as a sensitivity diagnostic. 

The accuracy of the fitted dark-matter cell masses can be quantified by the maximum fractional residual
\begin{equation}
\epsilon_{\rho,\max}
=
\max_{j\in{\rm fit}}
\left|
\frac{\sum_i p_{ij}a_i-b_j}{b_j}
\right|.
\label{eq:schwarzschild-density-residual}
\end{equation}
For the $8\times10^5$-orbit reconstructions, the Model~1 spherical fit has $\epsilon_{\rho,\max}=2.4\times10^{-6}$. The axisymmetric fits for Models~2, 3, and 4 have maximum constrained-cell fractional residuals of $2.9\times10^{-5}$, $2.5\times10^{-5}$, and $2.8\times10^{-5}$, respectively. The $1.6\times10^6$-orbit Model~4 fit gives $2.1\times10^{-5}$. These values apply to the fitted spatial constraints, not the deliberately excluded low-mass cells. 

For the axisymmetric models, we extract the local dark-matter velocity distribution from crossings of the Galactic midplane within a narrow radial interval around $R_0=8\,\kpc$. Because the density constraints are axisymmetric and each cylindrical cell spans the full azimuthal range, we integrate over $\phi$ from the outset. Starting from the cylindrical volume element $d^3x=R\,dR\,d\phi\,dz$, integration over $0\leq\phi<2\pi$ gives
\begin{equation}
dV=2\pi R\,dR\,dz.
\label{eq:axisymmetric-volume-element}
\end{equation}
For a point at $(R_i(t),z_i(t))$, the corresponding axisymmetric delta-function representation is
\begin{equation}
\frac{
\delta[R-R_i(t)]\,
\delta[z-z_i(t)]
}{2\pi R}.
\label{eq:axisymmetric-delta}
\end{equation}
It is normalized because
\begin{equation}
\int_0^\infty dR
\int_{-\infty}^{\infty} dz
\int_0^{2\pi} d\phi\,
R\,
\frac{
\delta[R-R_i(t)]\,
\delta[z-z_i(t)]
}{2\pi R}
=1.
\label{eq:axisymmetric-delta-normalization}
\end{equation}
The time-averaged density contribution of orbit $i$ is therefore
\begin{equation}
\rho_i(R,z)
=
\frac{w_i}{T_i}
\int_0^{T_i} dt\,
\frac{\delta[R-R_i(t)]\,\delta[z-z_i(t)]}{2\pi R},
\label{eq:orbit-density-axisymmetrized}
\end{equation}
where $T_i$ is the total integration time of orbit $i$ in the Schwarzschild library, defined by the $50T_{\rm circ}$ prescription above.

To connect this expression explicitly with the standard Schwarzschild cell weights, consider a cell $j$ in the $(R,|z|)$ grid. Integrating the density contribution of orbit $i$ over that cell gives
\begin{align}
\int_{\mathrm{cell}\ j}\rho_i\,dV
&=
\frac{w_i}{T_i}
\int_0^{T_i}dt
\int_{\mathrm{cell}\ j}
\frac{
\delta[R-R_i(t)]\,\delta[z-z_i(t)]
}{2\pi R}
2\pi R\,dR\,dz
\label{eq:cell-mass-integral}
\\
&=
\frac{w_i}{T_i}
\int_0^{T_i}dt\,
\mathbf{1}_j[R_i(t),z_i(t)]
\label{eq:cell-indicator-integral}
\\
&=
w_i\frac{\Delta t_{ij}}{T_i}
=
w_i p_{ij}.
\label{eq:cell-occupancy-mass}
\end{align}
Here $\mathbf{1}_j$ is unity while the orbit lies in cell $j$ and zero otherwise, $\Delta t_{ij}$ is the total time orbit $i$ spends in that cell, and
\begin{equation}
p_{ij}\equiv\frac{\Delta t_{ij}}{T_i}
\label{eq:cell-occupancy-fraction}
\end{equation}
is its cell-occupancy fraction. Thus the mass contributed to cell $j$ by orbit $i$ is its weight $w_i$ multiplied by the fraction of time $p_{ij}$ that it spends in the cell. The Schwarzschild weights are chosen so that $\sum_i w_i p_{ij}$ reproduces the target mass in each fitted cell~\cite{BinneyTremaine2008,VasilievAthanassoula2015}.

 In the azimuthally averaged orbit representation, we express velocities in the instantaneous local cylindrical basis, $\mathbf v=(v_R,v_\phi,v_z)$, with $\mathbf v_i(t)$ evaluated at each orbit's position. The resulting phase-space density is
\begin{equation}
g(R,z,\mathbf v)
=
\sum_i
\frac{w_i}{T_i}
\int_0^{T_i}dt\,
\frac{\delta[R-R_i(t)]\,\delta[z-z_i(t)]}
{2\pi R}\,
\delta^{(3)}\!\left[\mathbf v-\mathbf v_i(t)\right].
\label{eq:orbit-phase-space-density}
\end{equation}
For a finite orbit library, evaluating this distribution at exactly $R=R_0$ and $z=0$ is impractical. We therefore define a uniform-in-$R$ average of the midplane phase-space density over a narrow radial window,
\begin{equation}
\overline{g}_{\Delta R}(\mathbf v)
=
\int dR\,
K_{\Delta R}(R-R_0)\,
g(R,0,\mathbf v),
\label{eq:window-averaged-phase-space-density}
\end{equation}
where
\begin{equation}
K_{\Delta R}(R-R_0)
=
\begin{cases}
\dfrac{1}{2\Delta R}, & |R-R_0|<\Delta R,\\[6pt]
0, & \text{otherwise},
\end{cases}
\label{eq:radial-top-hat}
\end{equation}
and therefore
\begin{equation}
\int_{-\infty}^{\infty}
K_{\Delta R}(R-R_0)\,dR
=1.
\label{eq:radial-top-hat-normalization}
\end{equation}
Thus $\Delta R$ is the half-width of a radial sampling interval centered on $R_0=8\,\kpc$, whose full width is $2\Delta R$.
The corresponding normalized velocity distribution is
\begin{equation}
f_{\Delta R}(\mathbf v)
=
\frac{\overline{g}_{\Delta R}(\mathbf v)}
{\displaystyle\int d^3v\,\overline{g}_{\Delta R}(\mathbf v)}.
\label{eq:window-averaged-velocity-distribution}
\end{equation}

The vertical delta function can be evaluated exactly at the midplane using the standard identity for the Dirac delta function of a differentiable function $F(x)$ with isolated simple zeros $x_k$:
\begin{equation}
\delta[F(x)]
=
\sum_k
\frac{\delta(x-x_k)}{|F'(x_k)|}.
\label{eq:delta-function-composition}
\end{equation}
Applying this identity to $z_i(t)$, whose zeros $t_{ik}$ are the midplane-crossing times, gives
\begin{equation}
\delta[z_i(t)]
=
\sum_k
\frac{\delta(t-t_{ik})}{|v_{z,ik}|},
\label{eq:midplane-delta}
\end{equation}
where $v_{z,ik}=v_{z,i}(t_{ik})$. Substituting Eqs.~\eqref{eq:orbit-phase-space-density} and \eqref{eq:midplane-delta} into Eq.~\eqref{eq:window-averaged-phase-space-density} gives
\begin{equation}
\overline{g}_{\Delta R}(\mathbf v)
=
\sum_{i,k}
\frac{w_i\,K_{\Delta R}(R_{ik}-R_0)}
{T_i |v_{z,ik}|\,2\pi R_{ik}}
\delta^{(3)}\!\left(\mathbf v-\mathbf v_{ik}\right),
\label{eq:window-averaged-crossings-kernel}
\end{equation}
where $R_{ik}=R_i(t_{ik})$ and $\mathbf v_{ik}=\mathbf v_i(t_{ik})$ are the cylindrical radius and velocity of orbit $i$ at its $k$-th midplane crossing. For crossings within the accepted radial window, $K_{\Delta R}=1/(2\Delta R)$, so their unnormalized weights are
\begin{equation}
W_{ik}
\equiv
\frac{w_i}
{T_i |v_{z,ik}|\,2\pi R_{ik}\,2\Delta R}.
\label{eq:crossing-weight}
\end{equation}
The window-averaged phase-space density can therefore be written as
\begin{equation}
\overline{g}_{\Delta R}(\mathbf v)
=
\sum_{i,k}
W_{ik}
\delta^{(3)}\!\left(\mathbf v-\mathbf v_{ik}\right).
\label{eq:window-averaged-crossings}
\end{equation}
Here and in the equations below, the sums over $i,k$ include only accepted midplane crossings satisfying $|R_{ik}-R_0|<\Delta R$.
Integrating over velocity and using the normalization of the three-dimensional Dirac delta function gives
\begin{equation}
\int d^3v\,\overline{g}_{\Delta R}(\mathbf v)
=
\sum_{i,k}W_{ik}.
\label{eq:crossing-velocity-normalization}
\end{equation}
Substituting Eqs.~\eqref{eq:window-averaged-crossings} and \eqref{eq:crossing-velocity-normalization} into Eq.~\eqref{eq:window-averaged-velocity-distribution} gives the raw weighted empirical distribution of midplane-crossing velocities
\begin{equation}
f_{\Delta R}(\mathbf v)
=
\frac{1}{\sum_{i,k} W_{ik}}
\sum_{i,k}
W_{ik}\,
\delta^{(3)}\!\left(\mathbf v-\mathbf v_{ik}\right),
\label{eq:local-velocity-empirical}
\end{equation}
This raw distribution retains the velocity components of each accepted crossing. For the axisymmetric initial conditions, we instead sample the distribution obtained by independently randomizing the signs of $v_\phi$ and $v_z$:
\begin{equation}
\begin{aligned}
f_{\Delta R}^{\rm samp}(\mathbf v)
={}&
\frac{1}{4\sum_{i,k}W_{ik}}
\sum_{i,k}W_{ik}
\sum_{s_\phi=\pm1}\sum_{s_z=\pm1}\\
&\times
\delta^{(3)}\!\left[
\mathbf v-
\left(v_{R,ik},s_\phi v_{\phi,ik},s_z v_{z,ik}\right)
\right].
\end{aligned}
\label{eq:local-velocity-symmetrized}
\end{equation}
Each combination of $s_\phi$ and $s_z$ has probability $1/4$. This distribution is symmetric under $v_\phi\mapsto-v_\phi$ and $v_z\mapsto-v_z$.
We adopt $\Delta R=0.05\,\kpc$ for axisymmetric velocity sampling. We assess sensitivity to the sampling window using half-widths of $0.05$, $0.10$, and $0.20\,\kpc$. For each pair of adjacent windows, we require the cumulative distributions of the speed to differ by less than $0.02$, those of each velocity component to differ by less than $0.03$, and each velocity dispersion to change by less than $3\%$. The $8\times10^5$-orbit libraries pass these criteria. For the spherically symmetric Model~1, we instead evaluate the conditional velocity distribution at exactly $r=8\,\kpc$.

\subsection{Minihalo disruption calculation}

We assume $m_a=25\,\upmu\mathrm{eV}$ throughout and quantify survival by the weighted cumulative mass fraction above $M_{\rm cut}=10^{-12}\,\Msun$,
\begin{equation}
f_{\rm surv}(M_{\rm cut})
=
\frac{\rho_{\rm surviving}(M_{\rm rem}\geq M_{\rm cut})}
{\rho_{\rm unperturbed}(M_{\rm init}\geq M_{\rm cut})},
\label{eq:surviving-mass-fraction}
\end{equation}
where the numerator sums the surviving remnant masses of minihalos whose final mass \(M_{\rm rem}\) is at least \(M_{\rm cut}\), and the denominator sums the initial masses \(M_{\rm init}\) of the unperturbed population that are at least \(M_{\rm cut}\).

For Models~1--4 and both benchmarks, we use the same weighted grid of $1000$ infall-redshift values and $1000$ minihalo masses, following the mass--redshift weighting approach of DSouza et al.~\cite{DSouza2025}. The grid points are $z_i=151^{i/1000}-1$ ($i=0,\ldots,999$) and $M_j=10^{-14+11j/999}\,\Msun$ ($j=0,\ldots,999$). We assign one Galactic orbit to each of the $10^6$ grid cells.
The redshift assigned to each grid cell specifies the minihalo infall epoch and hence the lookback-time interval over which stellar encounters are accumulated. All Galactic orbits begin at $R=8\,\kpc$, $z=0$. We take this initial point to represent the present epoch and integrate forward for the required lookback-time interval.  For Model~1, we sample velocities at $r=8\,\kpc$ from the spherical Schwarzschild distribution described in Appendix~\ref{app:spherical-schwarzschild}. For the Eddington benchmark, we instead draw velocities at $r=8\,\kpc$ from the isotropic distribution in Eq.~\eqref{eq:eddington-local-velocity-distribution}. The orbit integration and disk stellar encounters are otherwise identical to Model~1.

For Models~2--4, we assign an initial velocity to each grid cell by independently selecting an accepted midplane crossing $(i,k)$, the $k$-th crossing of library orbit $i$, with probability
\begin{equation}
P_{ik}^{\mathrm{axi}}
=
\frac{W_{ik}}{\displaystyle\sum_{j,l}W_{jl}},
\label{eq:axisymmetric-velocity-selection-probability}
\end{equation}
where $W_{ik}$ is defined in Eq.~\eqref{eq:crossing-weight} and the sum runs over crossings in the radial window. Crossings are drawn with replacement, and only the velocity of the selected crossing is used. The minihalo trajectory itself starts at $R=8\,\kpc$, $z=0$. We then draw independent, equiprobable signs $s_\phi,s_z=\pm1$ and set the initial velocity to $\mathbf v=(v_{R,ik},s_\phi v_{\phi,ik},s_zv_{z,ik})$, thereby sampling Eq.~\eqref{eq:local-velocity-symmetrized}. The reflections of $v_\phi$ and $v_z$ leave the disruption histories unchanged in the adopted axisymmetric, midplane-symmetric gravitational potential and stellar-density model.

 Each Schwarzschild model draws its own Galactic initial velocities from its fitted local distribution. We integrate the trajectories with AGAMA and evaluate encounters with the hybrid prescription for combining multiple encounters of DSouza et al.~\cite{DSouza2025}.

In Models~1--4 and both benchmarks, the Galactic potential and the stellar-density profiles relevant to each calculation are held fixed over the minihalo lookback interval, rather than evolved with the assembly of the Milky Way. Shen et al.~\cite{Shen2024} explored two approximate time-dependent model variations separately (their Appendix E). Allowing the stellar disk surface density to decrease at earlier times increased minihalo survival, whereas modifying minihalo orbital energies to account for growth of the host halo decreased survival. These effects had opposite signs and comparable magnitudes in their model, suggesting that they might partly offset each other. 

We also omit a separate prescription for stripping by the smooth Galactic tidal field and the associated evolution of minihalo internal structure. The mass loss calculated here arises from encounters with individual stars. Shen et al.~\cite{Shen2024} included both stellar encounters and smooth tidal stripping and found stellar encounters to dominate in their minihalo models.

\section{Results}
\label{sec:results}

\subsection{Fiducial calculations and baryonic effects}

Table~\ref{tab:spherical-validation} lists the surviving fractions for the singular-isothermal Maxwellian and truncated-NFW Eddington benchmarks. The singular-isothermal result of $29.60\%$ reproduces the $29.6\%$ reported by DSouza et al.~\cite{DSouza2025}.

\begin{table}[htbp]
\centering
\caption{Surviving minihalo mass fractions for two independent spherical benchmark calculations. Both use AGAMA orbit integration, the same hybrid disruption prescription, $m_a=25\,\upmu\mathrm{eV}$, and a minihalo mass threshold of $M_{\rm cut}=10^{-12}\,\Msun$ for calculating the cumulative surviving mass fraction.}
\label{tab:spherical-validation}
\begin{tabular}{@{}llr@{}}
\toprule
Galactic halo model & Initial velocity sampling & $f_{\rm surv}$ \\
\midrule
Singular isothermal & Maxwellian & 29.60\% \\
Truncated NFW & Eddington & 33.20\% \\
\bottomrule
\end{tabular}
\end{table}

Table~\ref{tab:local-speeds} gives the mean and root-mean-square (RMS) dark-matter speeds at the solar radius. The first row is the spherical Eddington benchmark. Models~1--4 use the local Schwarzschild velocity distributions constructed with $8\times10^5$ orbits and $\lambda=0.1$. Model~1 is evaluated at $r=8\,\kpc$, while Models~2--4 are evaluated at $R=8\,\kpc$, $z=0$. Including disk and bulge gravity raises both speed measures.

\begin{table}[htbp]
\centering
\caption{Mean and RMS dark-matter speeds at the solar radius. The Eddington benchmark and Model~1 have the same spherical gravitational potential and disk stellar encounter prescription but different velocity-distribution constructions. Models~1--4 use the Schwarzschild velocity distributions defined in Table~\ref{tab:model-definitions}. Values are rounded to the nearest $\kms$.}
\label{tab:local-speeds}
\begin{tabular}{@{}lcc@{}}
\toprule
Case & $\langle v\rangle$ ($\kms$) & $v_{\rm rms}$ ($\kms$) \\
\midrule
Eddington benchmark & 214 & 233 \\
Model~1 & 214 & 233 \\
Model~2 & 281 & 300 \\
Model~3 & 292 & 309 \\
Model~4 & 282 & 300 \\
\bottomrule
\end{tabular}
\end{table}

Table~\ref{tab:survival} gives the surviving fractions for the independent Eddington benchmark and Models~1--4, with $8\times10^5$ Schwarzschild orbits used for each numbered model. For Models~1--4, the quoted $\pm$ values are the finite-library sensitivities from Table~\ref{tab:library-sensitivity}. Figure~\ref{fig:survival-convergence} shows the corresponding orbit-library comparisons. Model~1 gives $(33.23\pm0.05)\%$ survival, approximately $0.03$ percentage points above the Eddington result of $33.20\%$ and within the Model~1 finite-library sensitivity.

\begin{table}[htbp]
\centering
\caption{Surviving minihalo mass fractions above the nominal $10^{-12}\,\Msun$ threshold, evaluated at the corresponding discrete mass-bin edge. The $\pm$ values for Models~1--4 denote indicative finite-library sensitivities from Table~\ref{tab:library-sensitivity}. They are not formal $1\sigma$ uncertainties or estimates of the total error. No finite-library sensitivity applies to the separate Eddington benchmark. All numbered models use $8\times10^5$ Schwarzschild orbits and $\lambda=0.1$.}
\label{tab:survival}
\begin{tabular}{@{}lp{0.36\textwidth}p{0.23\textwidth}r@{}}
\toprule
Case & Gravitational potential & Stellar encounters & $f_{\rm surv}$ \\
\midrule
Eddington & Spherical tNFW & Disk & 33.20\% \\
Model~1 & Spherical tNFW & Disk & $(33.23\pm0.05)\%$ \\
Model~2 & tNFW + disk & Disk & $(28.31\pm0.08)\%$ \\
Model~3 & tNFW + disk + bulge & Disk + bulge & $(28.06\pm0.02)\%$ \\
Model~4 & tNFW + lighter disk + bulge & Lighter disk + bulge & $(29.70\pm0.12)\%$ \\
\bottomrule
\end{tabular}
\end{table}

Using the spherical Schwarzschild Model~1 as the baseline, adding disk gravity (Model~1 to Model~2) reduces survival by $(4.92\pm0.09)$ percentage points, while adding bulge gravity and bulge stellar encounters (Model~2 to Model~3) reduces survival by a further $(0.25\pm0.09)$ percentage points. The $\pm$ values are obtained by combining the finite-library sensitivities in quadrature as described in Sec.~\ref{sec:convergence}. Overall, the decrease from the spherical Schwarzschild Model~1 to Model~3 is $(5.17\pm0.05)$ percentage points, combining the Model~1 and Model~3 finite-library sensitivities in quadrature.

To test whether forward integration from the present-day initial conditions biases the reconstructed disruption histories, we repeated the full Model~3 calculation with all three components of the initial velocities reversed for the same $10^6$ minihalos. In the static Galactic potential, integrating these reversed velocities forward in time traces the original trajectories backward in time. The surviving mass fraction changes from $28.0604\%$ for forward integration to $28.0621\%$ for backward integration, a difference of $+0.0018$ percentage points at the adopted mass threshold. This paired test indicates that the direction of integration has a negligible effect on our results. 

\subsection{Orbit-library convergence and regularization}
\label{sec:convergence}

\begin{figure}[htbp]
\centering
\includegraphics[width=\textwidth]{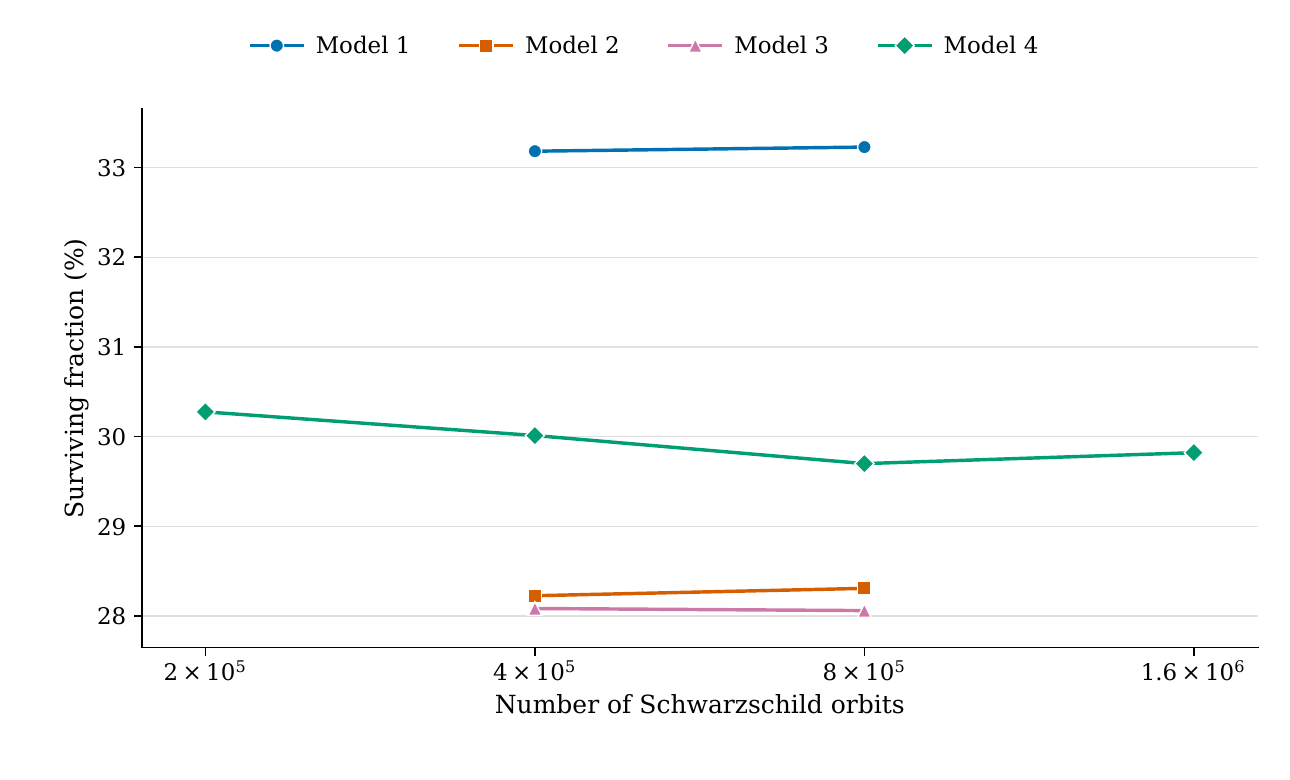}
\caption{Convergence of the surviving minihalo mass fraction with Schwarzschild library size for Models~1--4 (Table~\ref{tab:model-definitions}) at $\lambda=0.1$. Each point uses $10^6$ minihalos.}
\label{fig:survival-convergence}
\end{figure}

Figure~\ref{fig:survival-convergence} compares the fixed-$\lambda=0.1$ survival results for Models~1--4. Table~\ref{tab:library-sensitivity} gives the changes on increasing the Schwarzschild trial-orbit library from $4\times10^5$ to $8\times10^5$ for Models~1--3, and from $8\times10^5$ to $1.6\times10^6$ for Model~4. For Model~4, the change from $4\times10^5$ to $8\times10^5$ orbits was $-0.32$ percentage points, much larger than for the other models. We therefore ran a $1.6\times10^6$-orbit calculation, which tests the adopted $8\times10^5$ result directly. The absolute changes are used as simple estimates of finite-library sensitivity. For a difference between two Schwarzschild models, we combine their adopted sensitivities in quadrature, $\delta_{\rm lib}(f_A-f_B)=(\delta_{{\rm lib},A}^2+\delta_{{\rm lib},B}^2)^{1/2}$. As each calculation also uses independently sampled minihalo velocities, the measured library-size changes include Monte Carlo fluctuations. These estimates are not formal $1\sigma$ uncertainties and do not include uncertainty in the Galactic mass model.

\begin{table}[htbp]
\centering
\caption{Finite-library changes in the surviving mass fraction. The signed change $\Delta f_{\rm surv}$ is the larger-library result minus the smaller-library result. Its magnitude $\delta_{\rm lib}=|\Delta f_{\rm surv}|$ is used as the indicative numerical sensitivity in Table~\ref{tab:survival}. All changes are in percentage points (pp). Independent Monte Carlo fluctuations also contribute to them.}
\label{tab:library-sensitivity}
\begin{tabular}{@{}lcc@{}}
\toprule
Model & Trial-orbit comparison & $\Delta f_{\rm surv}$ (pp) \\
\midrule
1 & $4\times10^5\to8\times10^5$ & $+0.046$ \\
2 & $4\times10^5\to8\times10^5$ & $+0.082$ \\
3 & $4\times10^5\to8\times10^5$ & $-0.023$ \\
4 & $8\times10^5\to1.6\times10^6$ & $+0.122$ \\
\bottomrule
\end{tabular}
\end{table}

As a regularization diagnostic, we fit the same $4\times10^5$-orbit Model~3 library with both $\lambda=0$ and $0.1$. Because the radial sampling windows are narrow, the local velocity distribution should be nearly independent of their width, so we compare the local azimuthal velocity dispersion $\sigma_\phi$ evaluated in two windows as a consistency check. The relative difference between the intervals $|R-R_0|<0.20\,\kpc$ and $|R-R_0|<0.10\,\kpc$ decreases from $3.25\%$ for $\lambda=0$ to $2.54\%$ for $\lambda=0.1$, indicating that the regularized fit gives a more self-consistent local velocity distribution.  The corresponding surviving fractions, with both disk and bulge stellar encounters, are $28.12\%$ and $28.08\%$, a difference of $0.035$ percentage points calculated from the unrounded values. This is negligible compared with the $4.92$-percentage-point effect of disk gravity and comparable to the finite-library changes in Table~\ref{tab:library-sensitivity}.

\subsection{Sensitivity to the stellar disk mass}
\label{sec:disk-mass-sensitivity}

The fiducial stellar disk in Model~3 has a total mass of $5.76\times10^{10}\,\Msun$ and gives a circular speed of $249\,\kms$ at $R=8\,\kpc$. Reducing the disk mass to $5.00\times10^{10}\,\Msun$ in Model~4 gives a circular speed of $241\,\kms$, with the halo and bulge held fixed. This is closer to the local circular-speed estimates of McMillan~\cite{McMillan2017} and Reid et al.~\cite{Reid2019}. The disk stellar density is reduced by the same factor, $f_{\rm disk}\simeq0.868$, in both the gravitational potential and the encounter calculation. As shown in Table~\ref{tab:survival}, the surviving fraction increases from $(28.06\pm0.02)\%$ in Model~3 to $(29.70\pm0.12)\%$ in Model~4, a rise of $(1.64\pm0.12)$ percentage points in the $8\times10^5$-orbit comparison. The Galactic velocities are independently sampled in the two calculations.

\section{Conclusions}
\label{sec:conclusions}

For $m_a=25\,\upmu\mathrm{eV}$, the surviving minihalo mass fraction above the nominal $10^{-12}\,\Msun$ threshold decreases from $(33.23\pm0.05)\%$ for the spherical truncated-NFW Schwarzschild Model~1 to $(28.06\pm0.02)\%$ when the disk and bulge contribute to the Galactic potential and both stellar populations provide disruptive encounters. The reduction is approximately $(5.17\pm0.05)$ percentage points, using the finite-library sensitivities combined in quadrature. Adding the disk potential accounts for most of this difference, while the combined additional effect of bulge gravity and bulge encounters is smaller. This reduction is substantially smaller than the $23.9$-percentage-point reduction reported by DSouza et al.~\cite{DSouza2025} when replacing linear addition of stellar-encounter energy injections with their hybrid treatment of multiple encounters. 

The main Schwarzschild comparisons use $8\times10^5$ orbits and regularization $\lambda=0.1$. Model~1 gives $(33.23\pm0.05)\%$ survival in the spherical tNFW potential and agrees closely with Eddington inversion. Increasing the orbit libraries from $4\times10^5$ to $8\times10^5$ changes the survival fractions by less than $0.1$ percentage points for Models~1--3. Model~4 exhibits greater sensitivity, with $(29.70\pm0.12)\%$ survival at $8\times10^5$ orbits and $29.82\%$ at $1.6\times10^6$ orbits. 

Reducing the disk mass from $5.76\times10^{10}\,\Msun$ in Model~3 to $5.00\times10^{10}\,\Msun$ in Model~4 increases survival by approximately $(1.64\pm0.12)$ percentage points in the $8\times10^5$-orbit comparison. A fixed-library regularization test of Model~3 also shows that $\lambda=0.1$ can improve a local velocity-window diagnostic while changing the predicted surviving fraction by only about $0.035$ percentage points. Taken together, the close agreement with the independent spherical Eddington calculation, the small density-fit residuals, the local velocity-distribution checks, and the orbit-library and regularization tests support the numerical reliability of our main results, particularly the substantial effect of disk gravity. These checks do not establish comparable precision for the smaller bulge contribution or remove uncertainties associated with the assumed Galactic mass model. The $1.64$-percentage-point change from a $13\%$ reduction in disk mass shows that the predicted survival is sensitive to the Galactic mass model. In future work, we plan to fit the halo, disk, and bulge jointly to current Milky Way observations, such as the local circular speed and rotation curve.

Beyond the specific survival fractions reported here, this work demonstrates how Schwarzschild orbit superposition can be used to construct equilibrium dark-matter velocity distributions in Galactic potentials with nonspherical baryonic components. The same approach may be useful for other calculations that depend on the local dark-matter velocity distribution.

\section*{Acknowledgments}
OpenAI Codex was used to assist with the development and debugging of some of the Python code.
We thank Ian DSouza for providing the computational code and data underlying our earlier minihalo-disruption studies~\cite{DSouzaGordon2024,DSouza2025}, including publicly archived code~\cite{DSouzaCode2024}. We also thank him for comments on the manuscript. CG thanks John Forbes for helpful conversations.  MK was supported by NSF Grant No. 2412361 and NASA ATP Grant No. 80NSSC24K1226.

\appendix
\section{Spherical Schwarzschild construction}
\label{app:spherical-schwarzschild}

Model~1 uses a Schwarzschild orbit-superposition solution in the spherical truncated-NFW halo potential. We validate this baseline against the independent Eddington calculation in the same potential. The main $8\times10^5$ trial orbits are constrained by 30 radial density-projection coefficients together with the total-mass normalization. Their non-negative weights $w_i$ are determined by the density-fitting procedure described in the main text. The Eddington distribution is used to propose the initial conditions for the orbit library, while the local velocity distribution of the Schwarzschild model is determined by the fitted orbit weights.

For the spherical density constraints, we use AGAMA's \texttt{DensitySphHarm} target with $l_{\max}=m_{\max}=0$. The radial grid comprises 25 logarithmically spaced nodes $r_j=(0.5\,\kpc)\,3000^{j/24}$ for $j=0,\ldots,24$, supplemented by nodes at $r/\kpc=\{7.8,8.0,8.2,161\}$. The node at $161\,\kpc$ coincides with the halo truncation radius $r_{\rm cut}$, where the NFW density transitions to its exponentially truncated outer form. The resulting 29 distinct radial nodes define piecewise-linear radial basis projections rather than the top-hat cells used in the axisymmetric models. AGAMA returns 30 density-projection coefficients for this grid.

To sample the local velocity distribution at a fixed radius, each Schwarzschild orbit is weighted by the fraction of time it spends near that radius. Consider an orbit $i$ with weight $w_i$ and radial period $T_{r,i}$. In the static spherical potential, the radial period is constant along a given orbit, although different orbits generally have different radial periods. Provided that $r$ lies strictly between the orbit's pericenter and apocenter, the orbit crosses a thin shell of width $dr$ at $r$ twice per radial period, once inward and once outward. Denoting its radial velocity at $r$ by $v_{r,i}(r)$, the total time spent in the shell per radial period is $dt=2\,dr/|v_{r,i}(r)|$. Since $w_i$ is the mass represented by the orbit, the differential mass contribution is
\begin{equation}
dM_i(r)
=
\frac{2w_i}{T_{r,i}|v_{r,i}(r)|}\,dr.
\label{eq:spherical-shell-occupancy}
\end{equation}
At fixed $r$, the common factor $2\,dr$ cancels when normalizing the probabilities over the orbits that reach that radius. Thus the orbit-selection weight is proportional to
\begin{equation}
\frac{w_i}{T_{r,i}|v_{r,i}(r)|}.
\label{eq:spherical-orbit-selection-weight}
\end{equation}
At the solar radius $r_0=8\,\kpc$, the normalized probability of selecting orbit $i$ is
\begin{equation}
P_i^{\mathrm{sph}}
=
\frac{\displaystyle w_i/[T_{r,i}|v_{r,i}(r_0)|]}
{\displaystyle\sum_{j} w_j/[T_{r,j}|v_{r,j}(r_0)|]},
\label{eq:spherical-velocity-selection-probability}
\end{equation}
where the sum is restricted to orbits that reach $r_0$.

After selecting an orbit, we evaluate its radial and tangential speeds at $r_0$ from its energy and angular momentum. We assign equal probabilities to inward and outward radial motion and draw the orientation of the tangential velocity uniformly in the plane perpendicular to the radial direction. This preserves the spherical symmetry of the local velocity distribution while retaining the radial and tangential speeds implied by the fitted Schwarzschild weights. Model~1 draws its initial velocities for the disruption calculation from this fitted Schwarzschild distribution. The independent Eddington benchmark instead draws velocities directly from the isotropic distribution of Eq.~\eqref{eq:eddington-local-velocity-distribution}.

\end{document}